\documentclass[showpacs,aps, twocolumn]{revtex4}
\usepackage{epsfig}

\input{tcilatex}
\begin{document}

\title{Quantum Algorithm for Obtaining the Energy Spectrum of Molecular
Systems}
\author{Hefeng Wang, and Sabre Kais}
\affiliation{Department of Chemistry, Purdue University, West Lafayette, IN 47907}
\author{Al\'an Aspuru-Guzik,}
\affiliation{Department of Chemistry and Chemical Biology, Harvard University, Cambridge,
MA 02138}
\author{Mark R. Hoffmann}
\affiliation{Department of Chemistry, University of North Dakota, Grand Forks, ND 58202}

\begin{abstract}
Simulating a quantum system is more efficient on a quantum computer than on
a classical computer. The time required for solving the Schr\"odinger
equation to obtain molecular energies has been demonstrated to scale
polynomially with system size on a quantum computer, in contrast to the
well-known result of exponential scaling on a classical computer. In this
paper, we present a quantum algorithm to obtain the energy spectrum of
molecular systems based on the multi-configurational self-consistent
field~(MCSCF) wave function. By using a MCSCF wave function as the initial
guess, the excited states are accessible; Entire potential energy surfaces
of molecules can be studied more efficiently than if the simpler
Hartree-Fock guess was employed. We show that a small increase of the MCSCF
space can dramatically increase the success probability of the quantum
algorithm, even in regions of the potential energy surface that are far from
the equilibrium geometry. For the treatment of larger systems, a
multi-reference configuration interaction approach is suggested. We
demonstrate that such an algorithm can be used to obtain the energy spectrum
of the water molecule.
\end{abstract}

\maketitle

\section{introduction}

Since the discovery of a polynomial quantum algorithm for factorization~\cite%
{shor}, other quantum algorithms that provide exponential speedup over their
classical counterparts have been found. Examples in diverse areas include
the computation of approximations to the Jones polynomial~\cite{zeph} and
certain instances of the hidden subgroup problem~\cite{bacon}. Feynman
observed that simulating a quantum system might be more efficient on a
quantum computer than on a classical computer~\cite{Fey}. Further work by
others has born out this early suggestion~\cite{sl, zalka, abrams,
lidarwang, lev, aa, guzikdyn}. Although a quantum computer to carry out the
calculations that we propose is not currently experimentally realizable,
many recent developments in quantum information technology~\cite{SCCnot,
Blatt, Petta, white} continue to get closer towards the implementation of
such a device.

In quantum chemistry, where molecular quantum systems are simulated on a
classical computer, one is restricted to employ a finite basis to span the
formally infinite Hilbert space that would describe the electronic structure
of a molecular system. The full configuration interaction~(FCI) method~\cite%
{helgaker} diagonalizes the molecular Hamiltonian to provide solutions to
the electronic structure problem that are exact within this basis. FCI
scales exponentially with respect to the size of the molecular system
studied and therefore is restricted to the treatment of small diatomic and
triatomic systems~\cite{LJ}. Recently, a quantum algorithm for the solution
of the FCI problem in polynomial time was proposed by Aspuru-Guzik \textsl{%
et al.}~\cite{aa}. This algorithm employed the HF wave function as a
reference for further treatment of the correlation effects by the FCI
Hamiltonian on the quantum computer. The excited states of molecular systems
are difficult to resolve by employing the HF wave function as an initial
trial state. The main reason for this difficulty is due to the fact that
contributions from several configuration state functions~(CSF) must be
considered if one is seeking a reasonable overlap of the trial state with
the exact wave function. In the quantum chemical study of molecular systems,
people are often interested in computing molecular properties, such as the
energy of the ground state and a few low-lying excited states. i.e., in
study of the spectroscopic properties of molecules. In such cases, an FCI
calculation might become too expensive even for a quantum computer for some
large systems. A multi-reference configuration interaction~(MRCI)--truncated
CI--calculation based on an a multiconfigurational self-consistent
field~(MCSCF) wave function can sometimes provide results within chemical
accuracy, but with much less computational work than FCI due to the smaller
Hilbert space associated with the calculation. It is difficult to describe
various regions of molecular potential energy surfaces, sometimes even
qualitatively correct, by using a single reference determinant. Many
reference determinants or configuration state functions are often required
for the description of bond-dissociation regions.

In this paper, we suggest a quantum algorithm to obtain energy eigenvalues
of a MRCI wave function of a molecular system using the MCSCF wave function
as initial input to a quantum computer. We show that by improving the
quality of the trial wave function, the proposed algorithm yields
substantially higher success probabilities than by employing the HF wave
function. The use of a MCSCF wave function simultaneously reduces the amount
of quantum computing resources needed and extends the range of reliable
quantum computations to excited states and treacherous regions of the
potential energy surface. Simulating a chemical system with a quantum
computer requires the mapping of the Fock space of the MCSCF wave function
to the Hilbert space of the quantum bits~(qubits) of a quantum computer. We
introduce a more compact mapping technique for molecules by employing
symmetry properties. This approach reduces the computational resources for
representing the wave function on a quantum computer and avoids the
state-crossing problem.

The structure of this work is as follows. In Sec.~\ref{CIsec} we will review
the implementation of the FCI scheme on a quantum computer. Sec.~\ref%
{MCSCFsec} describes the properties of the MCSCF wave function. In Sec.~\ref%
{CIMCSCFsec} we describe a quantum algorithm for using MCSCF trial wave
functions in a FCI quantum algorithm. In Sec.~\ref{Resultssec} we discuss
numerical evidence for the feasibility of this scheme as applied to
calculations for the water molecule. We finalize with a conclusions section.

\section{implementation of CI scheme on a quantum computer}

\label{CIsec}

A closed quantum system in the non-relativistic limit can be described by
its Schr\"odinger equation~(atomic units are used),

\begin{equation}
i\frac{\partial \psi }{\partial t}=\hat{H}\psi .
\end{equation}%
Feit~\cite{MDF} and coworkers suggested a method to solve the Schr\"{o}%
dinger equation based on the spectral properties of the solutions to the
time-dependent Schr\"{o}dinger equation. Its solution can be expressed as a
linear superpositions of eigenfunctions of the Hamiltonian,
\begin{equation}
\psi (r,t)=\sum_{n}A_{n}u_{n}(r)exp(-iE_{n}t)
\end{equation}%
where the function $u_{n}(r)$ satisfies the equation $\hat{H}%
u_{n}=E_{n}u_{n} $. The method requires a numerical solution of $|\psi
(r,t)\rangle $ and the correlation function $P(t)$:
\begin{equation}
P(t)=\langle \psi (r,0)|\psi (r,t)\rangle =\int \psi ^{\ast }(r,0)\psi
(r,t)dr,
\end{equation}%
where $|\psi (r,0)\rangle $ is the wave function at $t=0$. $P(t)$ can then
be expressed as
\begin{equation}
P(t)=\sum_{n}|A_{n}|^{2}exp(-iE_{n}t),
\end{equation}%
which can be Fourier transformed to display the energy spectrum of the
system as a set of sharp local maxima at $E=E_{n}$.
\begin{equation}
P(E)=\sum_{n}|A_{n}|^{2}\delta (E-E_{n}).
\end{equation}

A scheme similar to the one proposed by Feit can be implemented on a quantum
computer. Abrams and Lloyd~\cite{abrams} suggested finding eigenvalues and
eigenvectors using a quantum phase estimation technique. Eigenfunctions of
the Hamiltonian are also eigenfunctions of the unitary time-evolution
operator, $U(t) = \exp (-i \hat{H} t)$, whose eigenvalues can be expressed
as a phase factor. A quantum Fourier transform~\cite{nc}~(QFT) is used to
retrieve the phase in a binary expansion and thus obtain the eigenenergy.
This scheme has been proposed to simulate quantum systems, especially
Fermion systems, on a quantum computer~\cite{al2,sl,sw,aa}. If the
Hamiltonian can be decomposed by means of a split-operator technique~\cite%
{MDF, zalka, lidarwang, guzikdyn}, the quantum computational cost is
polynomial, it can provide an exponential speed increase over its classical
counterpart.

The details of the algorithm proceeds as follows~\cite{abrams, tm}: First,
one must prepare two quantum registers, one is the index register composed
of $m$ qubits, which are used as control qubits and to perform a QFT
operation. Another register of $n$ qubits is the target register that is
used to represent the wave function of the system. The index register is
initially prepared in the zero state $\vert 0\rangle$. The quantum bits of
the index register are entangled with successive binary powers of the
unitary evolution operator on the target register. After the time-evolution
of the target register, the index register encodes an eigenvalue of the time
evolution operator $U$ of the target system as a phase represented in a
binary notation. By performing a QFT, the phase, and therefore the
eigenvalue of the system can be obtained.

The algorithm begins by initializing the quantum computer into the state:

\begin{equation}
|\Psi _{0}\rangle =|0\rangle |\psi \rangle .
\end{equation}%
Performing a $\pi /2$ rotation on each qubit in the index register results
on the state
\begin{equation}
|\Psi _{1}\rangle =\frac{1}{\sqrt{M}}\sum_{j=0}^{M-1}|j\rangle |\psi \rangle
,
\end{equation}%
where $M=2^{m}$. By performing a series of controlled-$U$ operations on this
state, it is transformed into:
\begin{equation}
|\Psi _{2}\rangle =\frac{1}{\sqrt{M}}\sum_{j=0}^{M-1}\hat{U}^{j}|j\rangle
|\psi \rangle .
\end{equation}%
The approximate vector $|\psi \rangle $ can be written as a sum of
eigenvectors of $U$,

\begin{equation}
|\psi \rangle =\sum_{k}c_{k}|\phi _{k}\rangle ,
\end{equation}%
where $k$ sums over the dimensionality of the target register. The
eigenvalue associated with $|\phi _{k}\rangle $ is $e^{i\phi _{k}}$, which
can be written as $e^{2\pi i\omega _{k}/M}$, where $\omega _{k}\in \lbrack
0,M)$. Using this fact, the state can be rewritten as:

\begin{equation}
|\Psi _{2}\rangle =\sum_{k}c_{k}|\phi _{k}\rangle \frac{1}{\sqrt{M}}%
\sum_{j=0}^{M-1}e^{2\pi ij\omega _{k}/M}|j\rangle .
\end{equation}

A QFT performed on the index qubits will reveal the phases $\omega _{k}$ and
thereby the eigenvalues. It requires $\sim m^{2}$ operations~\cite{nc}. A
polynomial number of trials are required to obtain any eigenvalue for which
the corresponding eigenvector is not exponentially small in the initial
guess. If the initial guess is close to the desired state, then only a few
trials may be necessary. Once a measurement is made and an eigenvalue is
determined, the target register qubits will collapse into the state of the
corresponding eigenvector.

Aspuru-Guzik \textsl{et al.}~\cite{aa} extended the algorithm to the study
of molecular systems and simplified the algorithm by introducing a recursive
phase-estimation technique that saves the qubits for performing the phase
estimation. They also introduced an adiabatic state preparation~(ASP)
technique for obtaining molecular ground states. They demonstrated that such
algorithms can be applied to problems of chemical interest using modest
numbers of quantum bits.

\section{MCSCF wave function}

\label{MCSCFsec}

In the previous quantum computing for quantum chemistry work~\cite{aa}, a HF
wave function was used as the initial trial wave function. The HF method~%
\cite{sbo} represents the wave function as a single Slater determinant. In
most cases, the HF wave function by itself is not sufficiently accurate to
generate useful chemical predictions such as relative energies of products
and reactants, and therefore a correlated calculation is necessary. Often,
the HF wave function is not a good initial guess to the exact wave function
of the system, especially for the excited states calculations, in which
contributions from several Slater determinants must be considered, even for
a qualitatively correct description. If a number of electron configurations
are relatively close in energy~(i.e. degenerate or near-degenerate), then
the HF approximation is particularly poor. This is the usual case when one
explores regions of avoided crossings~(or anti-crossings), molecules close
to the dissociation limit, in the limit of large system size, or in the
study of a chemical reaction path~\cite{gordon} In such cases, it is more
appropriate to describe the system with more appropriate wave functions in
which several different electron configurations are taken into account.

One realization of such a wave function comes from MCSCF theory. The general
form of an MCSCF wave function is:
\begin{equation}
\psi _{MCSCF}=\sum_{K}D_{K}\Phi _{K},
\end{equation}%
\begin{equation}
\Phi _{K}=(N!)^{-1/2}det|{\prod_{i\subset K}\phi _{i}}|,
\end{equation}%
\begin{equation}
\phi _{i}=\sum_{\mu }\chi _{\mu }C_{\mu i},
\end{equation}%
which is a linear combination of several electron configuration state
functions~(CSF). Each CSF differs in how the electrons are distributed
between the molecular orbitals~(MOs), $\phi _{i}$. For a particular system,
the CSFs can be chosen based on physical consideration of the system. The
MOs are usually expanded in a basis of atomic orbitals~(AOs), $\chi _{\mu }$%
. To obtain a MCSCF wave function, both the configuration expansion
coefficients $D_{K}$ and the MO expansion coefficients $C_{\mu i}$ are
variationally optimized. Hence the optimized vector is the best
approximation to the exact wave function of the system in a specific
parameter space. For a given set of orbital and configuration parameters,
even in a small variational space, the MCSCF wave function can give a much
better approximation than the HF wave function. A truncated CI based on an
MCSCF wave function, the so-called multi-reference CI method, normally, give
better results than a CI using a HF wave function as a reference, when small
Hilbert spaces are involved. The trade-off between the MRCI approach and the
FCI approach is that chemical intuition is involved in selecting the
appropriate CSFs for constructing the CI expansion.

As mentioned above, computational resource requirements are significantly
less for any reasonable MCSCF calculation than for an FCI calculation in the
same orbital space. Simple combinatorial arguments show that there are $%
\left(
\begin{array}{c}
2M \\
N%
\end{array}%
\right) $ possible Slater determinants formed from $M$ molecular orbitals
and $N$ electrons, and although their number can be reduced by space- and
spin-symmetry considerations, the growth in the number of determinants with
system size remains exponential. Even a well-constructed algorithm that uses
an iterative process for a subset of the roots~(such as that by Lanzcos or
Davidson)~\cite{ERD} will have CPU requirements that scale roughly as the
square of the number of determinants. Moreover, the storage requirements
scale as the number of determinants. Consequently, the FCI problem scales
exponentially with system size.

In contrast, MCSCF uses an iterative process to obtain an optimal~(in the
variational sense) space of specific size, e.g., $8$ electrons in $12$
orbitals. Even if one adopts the most costly~(but simply definable) MCSCF
calculation, the so-called Complete Active Space SCF~(CASSCF), and the
resulting number of determinants is given by the same formula as for full
CI, the CASSCF space is a tiny fraction of the full CI space. The number of
iterations required to determine the optimal space is usually on the order
of $10$. The non-CI part of an MCSCF calculation is typically dominated by
the transformation of electron repulsion integrals from the atomic basis in
which they are calculated to the molecular orbital basis, which scales as $%
M^5$. In fact, in real MCSCF calculations, the integral transformation step
is often the limiting step. It is worth noting that more complex MCSCF
calculations can be made possible by the use of the macroconfiguation
approach~\cite{YK}, which can reduce the number of determinants in an MCSCF
to a polynomial number even for larger orbital spaces. This essentially
guarantees that a physically meaningful and mathematically robust MCSCF
calculation will be integral bound, and therefore scale as $M^5$ with system
size~\cite{roos1, roos2} for systems of tens of atoms. Asymptotically, MCSCF
has an exponential cost as well and therefore a quantum computer still
provides an exponential speedup for this method.

In the MCSCF method, several states can be calculated simultaneously through
a state-averaged approach~\cite{wm,dy}. For the $n$-th MCSCF CI root, the
energy function can be written as,

\begin{equation}
E_{n}=\frac{\langle \psi _{n}|H|\psi _{n}\rangle }{\langle \psi _{n}|\psi
_{n}\rangle }.
\end{equation}%
A more general energy-like function can be constructed by use of weighting
vector~\cite{MRH},
\begin{equation}
E=\sum_{i}{w_{i}E_{i},}
\end{equation}%
where $w_{i}$ is the weight for state $i$. So, if we are interested in a few
evenly or non-evenly weighted states, the MO expansion coefficients are
optimized for all these states. By diagonalizing the one-particle density
matrix, we can obtain the occupation numbers in Fock space for each state.
This will be used as initial guess and map to the qubits on a quantum
computer in the quantum algorithm proposed in this work.

\section{Implementation of a general CI algorithm based on an MCSCF wave
function on a quantum computer}

\label{CIMCSCFsec}

The first step for the proposed simulation algorithm is to map the wave
function of the system to the state of the target register. In quantum
chemistry basis set methods, many-particle molecular wave functions are
represented in terms of a single-particle basis expanded in terms of atomic
orbitals and a many-particle basis expanded in terms of Slater determinants
or CSFs. In the direct mapping~\cite{aa}, each qubit represents the
fermionic occupation state of a particular atomic orbital. The Fock space of
the molecular system is mapped to the Hilbert space of the qubits. The
direct mapping has the advantage of yielding a simple Trotter expansion in
terms of a polynomial number of second-quantized Fermion operators.

The compact mapping considers the restriction of the multiplicity of the
system and reduces the number of qubits to represent the wave function to a
Hilbert space where all the quantum states of the target register correspond
to valid electronic configurations within a given spin symmetry. The
challenge of employing the compact mapping to general quantum systems is
that the representation of the time-evolution operator may involve a larger
number of non-local quantum gates.

Here we introduce a more compact mapping technique, which considers the
symmetry restriction of the molecules. The electronic states can be
categorized into different irreducible representation of their point group.
The subspace associated with a particular irreducible representation can be
mapped to the Hilbert space of the target register. This results in
considerable savings in the number of qubits required to represent the wave
function. Since there is no interaction between states that belong to
different irreducible representations, this mapping technique can aid in
solving certain cases of the state crossing problem~\cite{helgaker}.

For the proposed scheme, the wave function of the desired state is
implemented as the initial input to the phase estimation algorithm using the
MCSCF approach. The approximation to the exact wave function of the $i$-th
state $\vert\Psi_i\rangle$, is $\vert\Psi_{i}^{MCSCF}\rangle$. The
probability of observing the exact $i$-th state is $\vert\langle\Psi_i\vert%
\Psi_i^{MCSCF}\rangle\vert^2$. Since the MCSCF wave function provides a much
better approximation to the ground state wave function of the system than
does the HF wave function, and also provides a better description of excited
states than a Koopmans' theorem estimate~\cite{helgaker} from a HF wave
function, the probability of obtaining the correct energy of the system in
the phase estimation procedure is higher for MCSCF wave functions than for
HF wave functions.

The first step for the quantum algorithm involves the preparation of a MCSCF
calculation for $N$ states that are of interest for the system. The MCSCF
wave function for the state of interest are used as the initial guess for
the trial wave function:

\begin{equation}
|\Psi \rangle =|\psi _{n}^{0}\rangle ,
\end{equation}%
where $|\psi _{n}^{0}\rangle $ is the MCSCF wave function for the $n$-th
state. The next step is to map the MCSCF wave function for the $n$-th state
as the initial input to the quantum computer. This will be prepared using a
state-preparation algorithm. General state preparation is a hard problem,
but generally the MCSCF wave function contains a polynomial number of
non-zero terms in the Hilbert space, and therefore may be prepared
efficiently~\cite{Soklakov2006}. Feeding the MCSCF wave function into the
phase estimation algorithm as initial guess, the eigen-energies of the
corresponding CI state can be retrieved.

An MCSCF vector can be expanded as follows:
\begin{equation}
|\psi _{n}^{0}\rangle =\sum_{k}c_{k}|\psi _{k}\rangle ,
\end{equation}%
where $|\psi _{k}\rangle $ is the eigenvector of the CI matrix. $%
|c_{n}|^{2}=|\langle \psi _{n}|\psi _{n}^{0}\rangle |^{2}$ is the
probability of obtaining the eigenvector $|\psi _{n}\rangle $. A CI vector
for the $n$-th state can be written as:
\begin{equation}
|\psi _{n}\rangle =|\psi _{n}^{m}\rangle +|\psi _{n}^{p}\rangle =|\psi
_{n}^{0}\rangle +|\psi _{n}^{dev}\rangle +|\psi _{n}^{p}\rangle ,
\end{equation}%
where $|\psi _{n}^{m}\rangle $ is the part of the CI vector in the model
space, which is used to construct the MCSCF wave function; $|\psi
_{n}^{p}\rangle $ is the part of the CI vector in the space external to the
model space; $|\psi _{n}^{dev}\rangle $ is the deviation of MCSCF wave
function $|\psi _{n}^{0}\rangle $ from $|\psi _{n}^{m}\rangle $, the
projection of the CI vector in the model space. Then we have:
\begin{equation}
\langle \psi _{n}^{0}|\psi _{n}\rangle =1+\langle \psi _{n}^{0}|\psi
_{n}^{dev}\rangle +\langle \psi _{n}^{0}|\psi _{n}^{p}\rangle .
\end{equation}%
The vectors in model space and external space are orthogonal, $\langle \psi
_{n}^{p}|\psi _{n}^{0}\rangle =\langle \psi _{n}^{p}|\psi _{n}^{m}\rangle =0$%
. We can see that if the deviation vector goes to $0$, the overlap of the
MCSCF vector with the CI vector is one, the algorithm will be deterministic.

\section{Application to the water molecule}

\label{Resultssec}

\begin{table*}[tbp]
\caption{Results for the first singlet excited state of the water molecule
using the phase estimation algorithm. The MCSCF and HF wave function are
used as initial guesses. The FCI energy is -83.464130 a.u., the exact energy
for using MCSCF wave function is -83.449186 a.u. and for using HF wave
function is -83.443206 a.u.}%
\begin{tabular}{|c|c|c|}
\hline
Digits~(qubits) & Energy~(MCSCF) & Energy~(HF) \\ \hline
2 & $-83\pm 2.07\times 10^{1}$ & $-83\pm 2.07\times 10^{1}$ \\
8 & $-83.6386\pm 5.04\times 10^{-1}$ & $-83.6386\pm 5.04\times 10^{-1}$ \\
16 & $-83.4486\pm 1.28\times 10^{-3}$ & $-83.4435\pm 1.27\times 10^{-3}$ \\
24 & $-83.44919786\pm 4.95\times 10^{-6}$ & $-83.44318182\pm 4.95\times
10^{-6}$ \\ \hline
\end{tabular}%
\end{table*}

We have performed a quantum simulation for the ground state and the first
singlet excited state of the water molecule using the cc-pVDZ basis set~\cite%
{dunning}. For the ground state, considering the C$_{2V}$ symmetry of the
water molecule, the HF wave function of water is:

\begin{equation}
(1a_{1})^{2}(2a_{1})^{2}(1b_{2})^{2}(3a_{1})^{2}(1b_{1})^{2}.
\end{equation}

We consider a complete active space~(CAS) type MCSCF method: the first two $%
a_1$ orbitals are frozen, the active space consists of $3 a_1 - 6 a_1$
orbitals, $1b_1$, and $1 b_2$ and $2 b_2$ orbitals. The MRCI is performed
using the same model space but considering the single and double excitations
to the external space. The MCSCF space contains $152$ CSFs. The CI space
contains $13872$ CSFs, here $log_2^{13872}=13.76$, so $14$ qubits are
required to represent the CI wave function on a quantum computer. The
geometry used in the calculation is near the equilibrium geometry~($%
R_0=1.8435 a_0$ and $\angle HOH=110.57$). We varied both OH bonds from $0.5$
to $10$ times of the equilibrium distance simultaneously, keeping the $%
C_{2v} $ symmetry, $R=aR_0, a=0.5 - 10$. The success probability of the
quantum algorithm for using HF and MCSCF wave function as initial input $%
\vert\langle\Psi_i^{HF}\vert\Psi_i^{CI}\rangle\vert^2$ and $%
\vert\langle\Psi_i^{MCSCF}\vert\Psi_i^{CI}\rangle\vert^2$, are shown in Fig.
1. By following the stretch coordinate, we observe that the success
probability for using MCSCF wave function as initial guess is very high~($%
>0.9$) through the stretching, while the success probability for using HF
wave function as initial guess decreases very fast as the OH bond is
stretched. We can still obtain high probability of success by using just a
few CSFs instead of all $152$ CSFs in the MCSCF model space. In Fig. 2, we
show the success probability for both the ground state and excited states
using $6$ and $8$ CSFs respectively. With a relatively small number of CSFs
one can have a reasonable overlap with the desired state.
\begin{figure}[tbp]
\includegraphics[width=\columnwidth, clip]{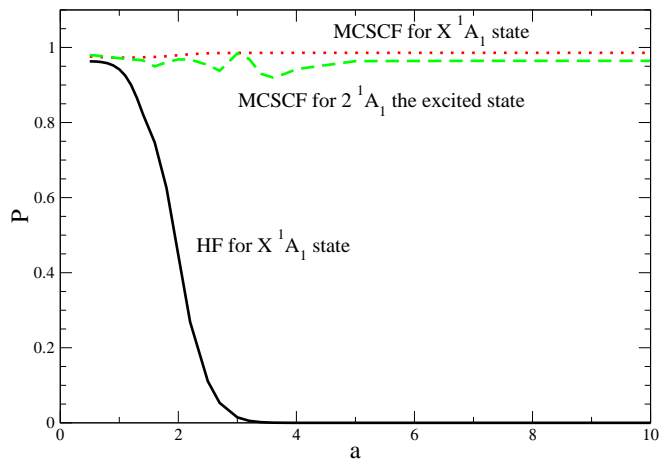}
\caption{Success probability ($P=|<\Psi^{HF}|\Psi^{CI}>|^2$ and $%
P=|<\Psi^{MCSCF}|\Psi^{CI}>|^2$) of using HF and MCSCF wave function as the
initial guess. Black line is for HF wave function, red line is for the MCSCF
wave function of the ground state, green line is the MCSCF wave function for
the excited state. The system is the water molecule, where $a=R/R_0$ is the
ratio between the stretched bond length $R$ and the bond length near
equilibrium distance $R_0=1.8435 a_0$. }
\end{figure}

\begin{figure}[tbp]
\includegraphics[width=\columnwidth, clip]{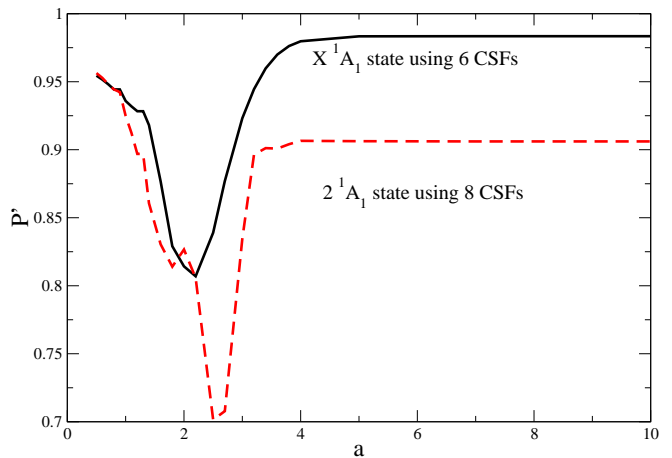}
\caption{ Success probability ($P^{\prime MCSCF}|\Psi^{CI}>|^2$)
of using a few CSFs as the initial guess. Black line is for the
ground state and red
line is for the excited state. The system is the water molecule, where $%
a=R/R_0$ is the ratio between the stretched bond length $R$ and
the bond length near equilibrium distance $R_0$. }
\end{figure}

We further studied the performance of the method for excited states. We
explored the first excited state of the water molecule at the equilibrium
geometry using the STO-3G basis set~\cite{sto}. The first two $a_1$ orbitals
were frozen. The model space for the MCSCF is a complete active space that
includes the $3a_1, 4a_1, 1b_1$ and $1b_2$ orbitals. The MRCI calculation
uses the same model space, but considers the single and double excitation to
the external space.

We use the scheme introduced by Parker and Plenio~\cite{PP} to implement the
QFT. This method is known as the measured quantum Fourier Transform~(mQFT)
approach. In this scheme, only one control qubit is used, more qubits are
saved for representing the wave function. The mQFT approach is based on the
fact that the gates within the Fourier transform are applied sequentially on
the qubits. Thus instead of performing the entire transform and then making
measurements on all control qubits afterwards, one can apply the single
qubit operation to the first qubit and then measure it. The operations
controlled by this first qubit are then replaced by single qubit operations
given the result of the measurement on the first. The measurement outcome is
fed back into the quantum calculation and this procedure is recycled till
all the required binary digits are resolved. The target register must remain
coherent during the whole procedure. For more details on this procedure, the
reader is referred to Ref.~\cite{PP}.

For the excited-state simulation, the CI space is composed of $18$ CSFs, so $%
5$ qubits are required to represent the wave function. In Table I, we
present the results for the calculation of the first excited state of the
water molecule in the STO-3G basis using the mQFT algorithm. The MCSCF wave
function and HF wave function are used as different initial guesses. The
MRCI energies are obtained to different digits of accuracy depending on the
number of ancillary control qubits employed. The error bars in the table
come from the numbers of the qubits in the index register. The more control
qubits in the index register, the more binary digits can be retrieved. For
example, if $n$ qubits are used as control qubits, then one can only obtain
up to $n$ binary digits of accuracy in the phase estimation, all the binary
digits after these $n$ digits will be uncertain. Therefore, the error is the
same regardless of the initial trial state~(HF or MCSF) employed. The FCI
energy is in this case is $-83.464130$ a.u. The first singlet excited state
energy for the water molecule using $24$ qubits~(E=-83.44919786 a.u.) is
lower, even including the error bars, than the exact energy using the MCSCF
wave function~(E=-83.449186 a.u.), this is because the error in expansion of
the unitary matrix is only up to the second order in Trotter expansion~\cite%
{aa}.

\section{discussion and conclusions}

In certain regions of molecular potential energy surfaces, electronic states
can cross each other or have low gaps, like in the case of avoided crossing
regions or at the bond-dissociation limit. In these cases, the interactions
between states should be considered simultaneously. Consequently, the use of
single determinant based methods is challenging.

Using an MCSCF wave function as the initial guess can deal with the strong
interaction between states straightforwardly. This can avoid possible
convergence of the state wave function to some undesired and unphysical
states when the energy gap between these states is small.

By using the more compact mapping technique, crossing states that belong to
different irreducible representations can be addressed separately since
there is no interaction between the states. For states in an avoided
crossing region and at the dissociation limit where states are near
degenerate, since the interaction has been considered qualitatively in the
MCSCF calculation, the overlap of the MCSCF wave functions with the
corresponding CI wave functions are still large, so that even in such
regions the probability for the reference states is high. Therefore, we
conclude that the MCSCF wave function can be used as a good initial guess
for correlated wave functions using quantum computing to explore the whole
potential energy surfaces for ground and excited states with high
probability of success.

Using an HF wave function as the initial guess chooses a path $\hat{H}^{HF}
\rightarrow \hat{H}$, for the evolution from the HF state to the CI state.
In our scheme, we choose the path $\hat{H}^{MCSCF} \rightarrow \hat{H}$, and
the states evolve from the MCSCF state to the MRCI state. Unlike in the case
of HF wave function in which the evolution is started from a single element
of the CI matrix, the MCSCF wave function starts the evolution from a small
matrix. This makes the evolution safer and faster, especially for a MRCI
space. From the simulation we can see that by including a few CSFs in the
initial guess, the success probability can be increased from very small to
near unity. This idea might be used in developing other quantum algorithms.

\section{Acknowledgment}

We would like to thank Brian Austin for critical reading of the paper, Wanyi
Jiang, Peter~J. Love and Ivan Kassal for helpful discussions. We acknowledge
the support of Purdue Research Foundation and the Harvard Faculty of Arts
and Sciences for their financial support. M.~R. H. acknowledges the National
Science Foundation~(NSF-0313907) for support of his contribution to the
research presented herein. A.~A. G. thanks the Army Research
Office~(ARO-52944-PH) for their support for conducting this research.

\end{document}